\documentclass[11pt,a4paper]{article}
\usepackage{jcappub}

\usepackage{graphicx}

\newcommand{\bk}{{\boldsymbol k}}
\newcommand{\bq}{{\boldsymbol q}}

\newcommand{\bx}{{\boldsymbol x}}
\newcommand{\bB}{{\boldsymbol B}}

\newcommand{\cA}{{\cal A}}
\newcommand{\cB}{{\cal B}}

\def\ave#1{\left\langle #1 \right\rangle}
\def\re{{\rm e}}

\def\gw{{\rm gw}}
\def\ini{{\rm in}}
\def\fin{{\rm fin}}

\title{Gravitational waves from the evolution of magnetic field after electroweak epoch}

\author[a,b]{Oleksandr Tomalak}
\author[c,b]{and Yuri Shtanov}

\affiliation[a]{Cluster of Excellence PRISMA, Institut f\"ur Kernphysik, Johannes Gutenberg Universit\"at,\\ Mainz, Germany} %
\affiliation[b]{Department of Physics, Taras Shevchenko National University,\\ Kiev 03022, Ukraine} %
\affiliation[c]{Bogolyubov Institute for Theoretical Physics,\\ Kiev 03680, Ukraine} %

\emailAdd{tomalak@uni-mainz.de}
\emailAdd{shtanov@bitp.kiev.ua}

\abstract{It was recently demonstrated that the evolution of helical magnetic field in the
primordial plasma at temperatures $T\gtrsim10$~MeV is affected by the phenomenon of
chiral quantum anomaly in the electroweak model, leading to a possibility of
self-sustained existence of magnetic field and chiral asymmetry in the electronic
distribution. This may serve as a mechanism for generating primordial magnetic field in
the early universe. Violent magnetic-field generation may lead to production of
gravitational waves which, regardless of the fate of magnetic field itself, survive until
today.  We estimate the threshold value of the initial chiral asymmetry above which the
generated gravitational waves would affect the big-bang nucleosynthesis and would show up
in the current and future experiments on gravitational-wave detection.}


\begin{document}

\maketitle \flushbottom

%\pacs{98.80.Cq, 98.70.Vc, 95.85.Sz}

\section{Introduction}

Observations have established the omnipresence of magnetic field in the universe of
various magnitudes and on various spatial scales.  Galaxies such as Milky Wave possess
regular magnetic fields of the order of $\mu$G, and coherent fields of the order of
$100~\mu$G are detected in distant galaxies \cite{Bernet:2008qp,Wolfe:2008nk}. Recently,
an evidence was obtained for the presence of magnetic field in intergalactic medium,
including voids \cite{Tavecchio:2010mk,Ando:2010rb,Neronov:1900zz}, with strengths
exceeding $\sim 10^{-15}$~G\@.  This supports the idea of cosmological origin of magnetic
fields, which are subsequently amplified in galaxies, probably by the dynamo mechanism
(see a review in \cite{Widrow:2002ud}).

The origin of cosmological magnetic field is a problem yet to be solved, while there
exist several mechanisms by which this could be achieved. They can broadly be classified
into inflationary and post-inflationary scenarios.  Both types still face problems to
overcome: inflationary magnetic fields usually turn out to be rather weak, while those
produced after inflation typically have too small coherence lengths (see
\cite{Widrow:2002ud,Kandus:2010nw} for a review of these mechanisms and assessment of
these difficulties).  There is a hope that these problems can be solved by taking into
account some additional mechanisms; for instance, the coherence length can be increased
by the so-called `inverse cascade' in turbulent hydrodynamics, which transfers power and
energy from short to long spatial scales.

One of the mechanisms of generation of cosmological magnetic fields which is currently
under scrutiny is based on the chiral (axial) quantum anomaly present in the theory of
electroweak interactions \cite{Joyce:1997uy}. If the numbers of right-handed and
left-handed electrons in the early hot universe happen to deviate from their equilibrium
values (for example, due to leptogenesis involving physics beyond the standard model), so
that the corresponding non-equilibrium effective chemical potentials $\mu_{\rm L}$ and
$\mu_{\rm R}$ differ from each other, then a specific instability arises with respect to
generation of helical (hypercharge) magnetic field. The generated helical magnetic field,
in turn, is capable of supporting the electron chiral asymmetry, thus prolonging its own
existence to cosmological temperatures as low as tens of MeV \cite{Boyarsky:2011uy}.

The processes of magnetic-field generation such as the one just described will be
accompanied by production of gravitational waves.  One of the consistency checks for such
scenarios is provided by the experimental upper bounds on the gravitational-wave
background. In this paper, we would like to address this issue to determine the
gravitational-wave background and to see whether it places any constraints on the
mechanism discussed above.

We consider magnetic field generated in the early hot universe on electroweak temperature
scale $T_\ini \sim 100$~GeV.  As mentioned above, such a field can be sustained by
quantum anomaly until much lower temperatures of the order of 10~MeV, with power being
permanently transformed from short to long spatial scales (the so-called `inverse
cascade').  It turns out that the whole process of magnetic-field evolution leads to
gravitational-wave production which is very sensitive to the characteristics of the
initial spectrum of the magnetic field, in particular, exponentially depends on the
magnitude of the initial chiral asymmetry.  One of the aims of this paper is to determine
the threshold for this latter quantity above which the generated gravitational waves
would affect the big-bang nucleosynthesis and would be detectable in the current and
future experiments on gravitational-wave detection.  We will see that such a threshold
for the conformal difference of chemical potentials $\Delta \mu_c = a (\mu_{\rm L} -
\mu_{\rm R})$ at $T_\ini \simeq 100$\,GeV is of the order $\Delta \mu_0 \sim 10^{-4}$.

The paper is organized as follows.  In Sec.~\ref{sec:mag}, we describe the spectra of the
helical magnetic fields; in Sec.~\ref{sec:gw-field}, we introduce the basic parameter of
the theory under consideration and present the main equations describing the generation
of gravitational waves by evolving magnetic fields; in Sec.~\ref{sec:constraints}, we
give the constraints on the initial energy density in magnetic field, stemming from
non-detection of the produced gravitational waves or from the theory of big-bang
nucleosynthesis and recombination, under the assumption of the initial value of chiral
asymmetry $\Delta \mu_0 = 6 \times 10^{-5}$, which was adopted in \cite{Boyarsky:2011uy},
and demonstrate that the level of the produced gravitational radiation is practically
negligible in this case; in Sec.~\ref{sec:chirality}, we determine the threshold of the
quantity $\Delta \mu_0$ (which turns out to be of the order of $10^{-4}$), above which
the generated gravitational waves would affect the big-bang nucleosynthesis and would be
detectable in the current and future experiments on gravitational waves; in
Sec~\ref{sec:summary}, we formulate our conclusions. Technical details on the theory of
generation of gravitational waves are presented in Appendix~\ref{sec:gwaves}; the power
spectrum of the generated gravitational waves is calculated in
Appendix~\ref{sec:gravcalc}; the simple case of a monochromatic magnetic field is
considered in Appendix~\ref{sec:monosource}.

\section{Helical magnetic fields}
\label{sec:mag}

The source of gravitational waves is the stress tensor of matter. In the flat space-time,
the stress tensor of dissipationless magnetohydrodynamics (MHD) is given by \cite{llvol8}
\begin{equation} \label{set0}
T_{ij} = \rho v_i v_j + p \delta_{ij} - \left( B_i B_j - \frac12 \delta_{ij} B^2 \right)
\, ,
\end{equation}
where $v_i$ is the fluid velocity, $p$ is pressure, and $B_i$ is the magnetic field.
During the epoch of gravitational-wave production under consideration in this paper, the
electric conductivity $\sigma$ of the cosmological plasma is very large, so that electric
field can be neglected. Only the fluid velocity and magnetic field contribute to the
traceless part, relevant to the production of gravitational waves (see
Appendix~\ref{sec:gwaves}).

In this paper, we concentrate on the magnetic part of the stress tensor, which we assume
to be the major contribution to the gravitational-wave production. The role of turbulence
in the process of magnetic field generation was under consideration, e.g., in
\cite{arXiv:0909.0622,Tashiro:2012mf}.

In an expanding universe, it is convenient to work in the comoving conformal coordinate
system $(\tau, \bx)$, in which the magnetic part of the stress tensor (\ref{set0}) takes
the form
\begin{equation} \label{set}
T_{i j} = - \frac{1}{a^2}  \left( B_i B_j - \frac12 \delta_{ij} B^2 \right)  \, ,
\end{equation}
where $a (\tau)$ is the scale factor.\footnote{The spatial indices are then always
raised, lowered, and contracted by using the Kronecker delta-symbol, so that, for
example, $B^2 = \delta^{ij} B_i B_j$. The components $B_i$ in (\ref{set}) coincide with
the components of the so-called comoving magnetic field, which is connected with the
observable magnetic field strength $\bB_{\rm obs}$ by the relation $\bB = a^2 \bB_{\rm
obs}$.}

A generic two-point correlation function for a divergence-free statistically homogeneous
and isotropic magnetic field in Fourier representation has the form \cite{Caprini:2003vc}
\begin{equation} \label{correl}
\ave{B_i (\bk) B_j^* (\bk')} = \left( 2 \pi \right)^3 \delta \left(\bk - \bk' \right)
\left[ P_{ij} (\bk) S(k) + i \epsilon_{ijs} \hat k^s A(k) \right] \, ,
\end{equation}
where $\hat k^i = k^i / k$, $ P_{ij} = \delta_{ij} - \hat{k}_i \hat{k}_j$ is the
symmetric projector to the plane orthogonal to $\bk$, and $ \epsilon_{ijk} $ is the
normalized totally antisymmetric tensor.

It is useful to introduce the helicity components $B_\pm (\bk)$ of the magnetic field via
\begin{equation}
B_i (\bk) = B_+ (\bk) \re^+_i (\bk) + B_- (\bk) \re^-_i (\bk) \, ,
\end{equation}
where the basis $ \re^\pm_i = \frac{1}{\sqrt{2}}\left( \re^1_i \pm i \re^2_i \right)$ is
formed from a right-handed orthonormal (with respect to the metric $\delta_{ij}$) basis
${\bf e}^{1} (\bk)$, ${\bf e}^{2} (\bk)$, ${\bf e}^{3} (\bk) = \bk /k$.  The symmetric
and helical parts of the correlation function are then expressed through these components
as follows:
\begin{eqnarray} \label{S}
S (k) &=& \frac12 \ave{|B_- (\bk)|^2 + |B_+ (\bk)|^2} \, , \\
\label{A} A (k) &=& \frac12 \ave{|B_- (\bk)|^2 - |B_+ (\bk)|^2} \, .
\end{eqnarray}
We note an obvious constraint $|A (k)| \leq S (k)$.

The helical part $A (k)$ of the magnetic-field correlation function characterizes the
difference in the power between the left-handed and right-handed magnetic field. The
symmetric part $S (k)$ characterizes the magnetic field energy density. Magnetic field
can be dominated by its left-handed or right-handed part. In this case, of the so-called
maximally helical magnetic field, one has $ |A(k)| = S(k)$.

Of relevance to the investigation in the present paper will be the case where the
right-handed and left-handed magnetic-field components evolve separately as
\begin{equation} \label{maggrowth}
B_\pm (\bk, \tau) = B_\pm (\bk) g_\pm (k, \tau) \, ,
\end{equation}
where $B_\pm (\bk)$ are the initial values of the field components at some moment of
time, and $g_\pm (k, \tau)$ are the corresponding growth factors. In this case, we can
work in the random-phase approximation for the initial field, thinking of the initial
amplitudes $|B_\pm (\bk)|$ as specified by the spectral functions $S (k)$ and $A (k)$ via
(\ref{S}) and (\ref{A}), and of their phases as of random and independent for different
$\bk$. In this case, the definitions of the correlation function (\ref{correl}) and
relations (\ref{S}) and (\ref{A}) are preserved with time, and the spectral functions
evolve as follows:
\begin{equation} \begin{array}{l}
S (k, \tau) = S (k) f_+ (k, \tau) + A (k) f_- (k, \tau)  \, , \medskip \\
A (k, \tau) = A (k) f_+ (k, \tau) + S (k) f_- (k, \tau)  \, .
\end{array}
\end{equation}
where
\begin{equation}
f_\pm (k, \tau) = \frac{|g_- (k, \tau)|^2 \pm |g_+ (k, \tau)|^2 }{2} \, .
\end{equation}
This ``random-phase'' approximation allows us to use the Wick's theorem for the
four-point correlation function.

If the mechanism of generation of magnetic field is local and causal, as will be assumed
in the present paper, then the magnetic-field power tends to zero as $k \to 0$ due to the
analyticity requirements. Indeed, the Taylor expansion of the spectrum at small $k$ then
has the form $S(k) \sim k^n$ with the condition $ n \geq 2$ stemming from the property of
the correlation function having compact support \cite{Caprini:2003vc}.

\section{Gravitational waves from helical magnetic fields}
\label{sec:gw-field}

According to the scenario proposed and described in \cite{Joyce:1997uy, Boyarsky:2011uy},
a configuration of maximally helical (hypercharge) magnetic field rapidly develops before
the electroweak phase transition in the presence of the electron chiral asymmetry due to
the effect of axial anomaly. Supported by the electron chiral asymmetry, magnetic field
can survive down to cosmological temperatures of the order of 10~MeV, after which the
effects of chirality flips and plasma conductivity become more efficient.

Depending on the sign of leptonic chiral asymmetry, one of the helicity components of
magnetic field gets amplified due to a specific instability, while the opposite helicity
component gets suppressed and can be neglected.  The correlation function of a maximally
helical magnetic field is then characterized by a single quantity $S (k)$ by virtue of
(\ref{S}) and (\ref{A}).  For definiteness, we will assume that the amplified component
is left-handed.

The evolution factor for the left-handed magnetic field in (\ref{maggrowth}) is given by
the expression \cite{Boyarsky:2011uy}
\begin{equation} \label{instab}
g_- (\bk , \tau ) \equiv g_k (\tau) = e^{- \cA (\tau) k^2 + \cB (\tau) k} \, ,
\end{equation}
with the following dimensionless coefficients of the powers of $k$:
\begin{equation} \label{AB}
\cA (\tau) =  \frac{\tau - \tau_\ini}{\sigma_c} \, , \qquad \cB(\tau) = \frac{\alpha} {2
\pi \sigma_c } \int^{\tau}_{\tau_\ini} \Delta \mu_c (\tau') d \tau' \, .
\end{equation}
Here, $\sigma_c \equiv a \sigma = {\rm const}$ \cite{Baym:1997gq} characterizes the
plasma conductivity, $\alpha \approx 1/137$ is the fine structure constant, and $\Delta
\mu_c \equiv a \left(\mu_L - \mu_R \right)$ is the (conformal) difference between the
chemical potentials of the left-handed and right-handed charged leptons. This last
quantity, in the presence of a maximally helical magnetic field, evolves according to the
system of equations \cite{Joyce:1997uy, Boyarsky:2011uy}\footnote{Here and below, an
overdot denotes the derivative with respect to the conformal time $\tau$.}
\begin{eqnarray}
\Delta \dot \mu_c (\tau) &=& - \frac{c_\Delta \alpha}{\pi^2} \int_0^\infty \dot S (\tau,
k) k d k
- \Gamma_{\rm f} (\tau) \Delta \mu_c (\tau) \, , \label{dmu}\\
\dot S (\tau, k) &=& \left( - \frac{2 k^2}{\sigma_c} + \frac{\alpha k \Delta \mu_c}{\pi
\sigma_c} \right) S (\tau, k) \, , \label{dS}
\end{eqnarray}
where $c_\Delta$ is a numerical factor of order unity, and $\Gamma_{\rm f} (\tau)$ is the
(time-dependent) rate of chirality flipping due to scattering processes.

As a characteristic example, one can consider a scenario \cite{Boyarsky:2011uy} where,
due to the growth instability described by (\ref{instab}), the magnetic field on a short
time scale $\tau \sim \tau_\ini$ develops a maximally helical state with spectrum of the
form
\begin{equation}\label{specinit}
S(k) = S_0 \left( \frac{k}{k_0} \right)^2 e^{- k^2/k_0^2} \, ,
\end{equation}
which is analytic at $k = 0$ and has a cut-off at the scale $k_0$.  Then, neglecting the
last (flipping) term in (\ref{dmu}), we see that there is an initial stable point
\begin{equation} \label{mu0}
\Delta \mu_0 = \frac{3 \pi k_0}{\alpha} \, ,
\end{equation}
at which one initially has $\Delta \dot \mu_c = 0$.

With the flipping term in (\ref{dmu}) taken into account, a typical law of the evolution
of the difference between the chemical potentials can be approximated by the power law
\cite{Boyarsky:2011uy}
\begin{equation} \label{evolu}
\Delta \mu_c (\tau) = \Delta \mu_0 \left( \frac{\tau}{\tau_\ini} \right)^{- \beta} \, ,
\end{equation}
in which $\Delta \mu_0$ is given by (\ref{mu0}).  The evolution in the form (\ref{evolu})
proceeds until the temperature drops to about $T \sim 10$~MeV at the corresponding time
$\tau_\fin$, after which this quantity rapidly decays as a consequence of dissipation.
The spectral density of magnetic field during this period evolves according to
(\ref{instab}).

In a radiation-dominated early universe, it is convenient to choose the scale factor as
the inverse of the temperature, $a = 1/ T$, since the product $a T$ is constant as long
as the number of relativistic degrees of freedom $g_*$ in the universe remains constant.
With this choice, we have $a = \tau / M_*$, where $M_* = \left( 45 / 4 \pi^3 g_*
\right)^{1/2} M_{\rm P}$ is the effective Planck mass. Then the relative plasma
conductivity $\sigma_c = \sigma / T \simeq 70$ \cite{Baym:1997gq}, the initial conformal
time $\tau_\ini = M_* / T_\ini \simeq 7.3 \times 10^{15} \left(100~{\rm GeV} / T_\ini
\right)$, and the final conformal time $ \tau_\fin = M_* / T_\fin \simeq 1.5 \times
10^{19} \left( 50~{\rm MeV} / T_\fin \right)$ with $g_* \approx 100$.

If we take the quantities $S_0$ and $\Delta \mu_0$ as our basic parameters in the initial
spectrum of magnetic field, then the cutoff wave number $k_0$ in (\ref{specinit}) is
determined from the value of $\Delta \mu_0$ via (\ref{mu0}).  For instance, for a
characteristic value $\Delta \mu_0 = 6 \times 10^{-5}$, we have $k_0 = 4.6 \times
10^{-8}$. In this case, the exponent $\beta \simeq 0.35$ according to the numerical
simulations of \cite{Boyarsky:2011uy}.  All our estimates below will refer to the case of
$\Delta \mu_0 = 6 \times 10^{-5}$, $T_\ini \simeq 100$~GeV and $T_\fin \sim 50$~MeV.

The wave number $k_m$ that maximizes the value of $g_k (\tau)$ in (\ref{instab})
decreases monotonically from $ k_m = 0.75 k_0 $ at $\tau = \tau_\ini$ to $ k_m = 0.08 k_0
$ at $\tau = \tau_\fin$.  For our typical values $T_\ini \simeq 100$~GeV and $T_\fin
\simeq 50$~MeV, the exponent in $ g_{k_m} (\tau)$ increases monotonically from zero to
$4.2 \times 10^{15} k_0^2 \approx 9$ for $\Delta \mu_0 = 6 \times 10^{-5}$, leading to a
considerable amplification of the initial magnetic field at large spatial scales and to
an `inverse-cascade' reddening of its spectrum.

Due to our choice of the scale factor as the inverse temperature, the initial magnetic
energy density is given by the expression
\begin{equation}
\rho_B (\tau_\ini) = \frac{T_\ini^4}{2 \pi^2} \int_0^\infty S(k) k^2 d k = \frac{3 S_0
k_0^3}{16 \pi^{3/2}}  T_\ini^4 \, .
\end{equation}
It is convenient to relate this quantity to the radiation energy density by introducing
the dimensionless parameter
\begin{equation} \label{rb}
r_B \equiv \frac{\rho_B (\tau_\ini)}{\rho_r (\tau_\ini)} = \frac{30 \rho_B}{\pi^2 g_*
T_\ini^4}  = \frac{15}{\pi^4 g_*} \int_0^\infty S (k) k^2 d k = \frac{90 S_0 k_0^3}{16
\pi^{7/2} g_*}\, .
\end{equation}
When making estimates, we will always assume this parameter to be smaller than unity,
which simply means that the energy density of magnetic field does not exceed the energy
density of radiation in the hot universe.

The details of the derivation of the produced gravitational radiation in the scenario
under consideration are presented in Appendices \ref{sec:gwaves} and \ref{sec:gravcalc}.
The energy density of gravitational waves $\rho_\gw$ is proportional to the square of the
initial amplitude $S_0$ of the magnetic-field spectral density. Therefore, it is
convenient to introduce a normalized dimensionless quantity which is independent of this
amplitude:
\begin{equation} \label{upsilon}
\Upsilon_\gw = \frac{\rho_\gw}{r^2_B \rho_r} \left[ \frac{g_* }{g (\tau)} \right]^{1/3} =
\frac{\Omega_\gw (\tau)}{r^2_B \Omega_r (\tau)} \left[ \frac{g_*}{g (\tau)} \right]^{1/3}
\, ,
\end{equation}
where $\Omega_\gw (\tau)$ and $\Omega_r (\tau)$ are the time-dependent fractions of the
energy density in gravitational waves and in radiation, respectively, and $g (\tau)$ is
the effective number of the degrees of freedom in radiation, initially equal to $g_*$.
This quantity has a natural spectral decomposition
\begin{equation} \label{upsilon1}
\Upsilon_\gw = \int \Upsilon_q\, d \ln q \, ,
\end{equation}
where the quantity $\Upsilon_q$ determines the spectral power of gravitational waves per
logarithmic frequency interval. Note that the quantity $\Upsilon_q$ is defined only for
wave numbers $q$ inside the Hubble radius, i.e., for $q \tau > 1$.  The energy density of
free gravitational waves evolves with time as $a^{-4}$.  As a consequence of entropy
conservation, the energy density of radiation evolves as $g^{-1/3} a^{-4}$.  Therefore,
the quantity $\Upsilon_q$ remains constant in time.

The effect of the inverse-cascade modification of the magnetic-field power spectrum by
the evolution factor (\ref{instab}) on the production of gravitational waves is
calculated in Appendix~\ref{sec:gravcalc}.  It depends on the main parameters of the
evolution of magnetic field.  For sufficiently large values of $q$, namely, for
\begin{equation} \label{qmin}
q \tau_\ini \gtrsim \left( \frac{9 k_0^2 \tau_\ini}{16 \sigma_c} \right)^{1/(1 - 2
\beta)} \, ,
\end{equation}
the quantity $\Upsilon_q$ can be approximated by the analytic formula
\begin{eqnarray}\label{final}
\Upsilon_q &\simeq& \left(\frac{q}{k_0} \right)^3 \left[ \ln^2 \left( 1 + \frac{1}{q
\tau_\ini} \right) \left( 1 + \frac{q^2}{k_0^2} \right) \phantom{\left(
\frac{\tau_\fin}{\tau_\ini} \right)^{2 \beta}} \right. \nonumber\\
&&\left. {} + R \ln^2 \left( 1 + \frac{1}{q \tau_\fin} \right) \left( 1 + \left(
\frac{\tau_\fin}{\tau_\ini} \right)^{2 \beta} \frac{q^2}{k_0^2} \right) e^{- 2 \cA_\fin
q^2} \right] e^{-q^2/k_0^2} \, ,
\end{eqnarray}
where the quantity $R$ is given by (\ref{app:ration}):
\begin{equation} \label{ration}
R = \frac{6}{k_0} \sqrt{\frac{\sigma_c}{\tau_\fin}} \left(\frac{\tau_\ini}{\tau_\fin}
\right)^{2 + 6 \beta} \exp \left[{\frac{9 k_0^2 \tau_\fin}{4 (1 - \beta)^2 \sigma_c}
\left( \frac{\tau_\ini}{\tau_\fin} \right)^{2 \beta}} \right]  \, .
\end{equation}

For small values of $q$, opposite to (\ref{qmin}), it is approximated by (\ref{final})
with $R = 0$\,:
\begin{equation} \label{ufinal}
\Upsilon_q \simeq \left(\frac{q}{k_0} \right)^3 \left( 1 + \frac{q^2}{k_0^2} \right)
\ln^2 \left( 1 + \frac{1}{q \tau_\ini} \right) e^{-q^2/k_0^2} \, .
\end{equation}

\begin{figure}[htp]
\begin{center}%\centering
\includegraphics[width=.8\textwidth]{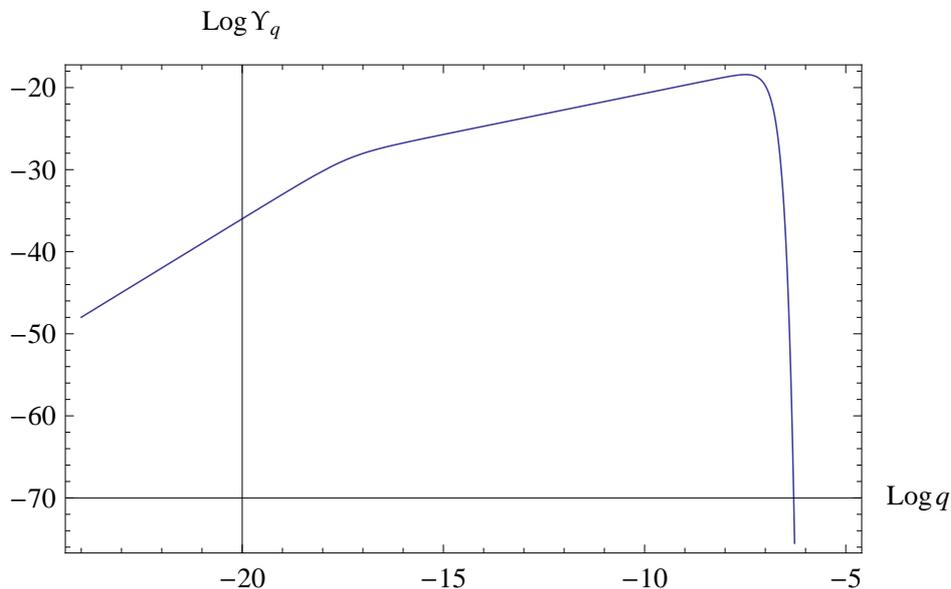}
\end{center}
\caption{The quantity $\log_{10} \Upsilon_q $ versus $ \log_{10} q$ as given by equation
(\ref{ufinal}) for the parameter values of the evolution of magnetic field described in
Sec.~\ref{sec:gw-field}. \label{fig:yq}}
\end{figure}

The term with $R$ in (\ref{final}) is shown in Appendix~\ref{sec:gravcalc} to be
insignificant under the characteristic model parameters $T_\ini \simeq 100~{\rm GeV}$,
$T_\fin \simeq 50~{\rm MeV}$, and $\Delta \mu_0 = 6 \times 10^{-5}$, studied in
\cite{Boyarsky:2011uy}. The quantity $\Upsilon_q$ in this case is approximated by
(\ref{ufinal}).  This function is plotted in Fig.~\ref{fig:yq} for the specified
evolution of magnetic field (\ref{instab})--(\ref{evolu}) with $k_0 = 4.6 \times
10^{-8}$, corresponding to $\Delta \mu_0 = 6 \times 10^{-5}$.  The spectrum of
gravitational waves rises at small wavenumbers and is exponentially suppressed at large
wavenumbers characteristic of the similar suppression in the magnetic-field power
spectrum, i.e., at $q \sim k_0$.

By integrating over all wave numbers, we find
\begin{equation} \label{utot}
\Upsilon_\gw \simeq \frac{1}{k_0^2 \tau_\ini^2} \simeq 10^{-17}\, ,
\end{equation}
where the numerical estimate is made for the parameter values in the scenario under
consideration.

\section{The background of gravitational waves}
\label{sec:constraints}

In this section, we calculate the background of gravitational waves depending on the
parameter $r_B$ of magnetic field defined in (\ref{rb}) and assuming the characteristic
model parameters $T_\ini \simeq 100~{\rm GeV}$, $T_\fin \simeq 50~{\rm MeV}$, $\Delta
\mu_0 = 6 \times 10^{-5}$, studied in \cite{Boyarsky:2011uy}.

To compare our calculations of the gravitational-wave background with various
constraints, we express our results in conventional notation.  The frequency $f$ of
gravitational waves today is related to its physical wavelength $\lambda_{\rm phys}$,
physical wave number $q_{\rm phys}$, and comoving wave number $q$ by the relations
\begin{equation} \label{qtonu}
f = \frac{1}{\lambda_{\rm phys}} = \frac{q_{\rm phys}}{2 \pi} = \frac{q T_0}{2\pi \hbar}
\left( \frac{2}{g_*} \right)^{1/3}\, ,
\end{equation}
where $ T_0 $ is the current temperature of the CMB.

The spectral energy density of the gravitational waves generated at a radiation-dominated
stage is expressed through $\Upsilon_q$ as
\begin{equation} \label{presday}
\Omega_\gw (f) \equiv \frac{8 \pi G}{3 H^2} \frac{d \rho_\gw}{d \ln f} = r^2_B \Omega_r
\left[ \frac{g (\tau)}{g_*} \right]^{1/3} \Upsilon_q \, .
\end{equation}
The quantities $q$ and $f$ in (\ref{presday}) are related with each other through
Eq.~(\ref{qtonu}).

Note that the comoving wave number associated with the Hubble radius for a
radiation-dominated universe ($ \ell_H = 1/H$) is given by
\begin{equation} \label{qH}
q_H = \frac{2 \pi a}{\ell_H} = \frac{2 \pi}{\ell_H T} \left[ \frac{g_*}{g(\tau)}
\right]^{1/3} = \frac{2 \pi T}{M_*} \left[ \frac{g(\tau)}{g_*} \right]^{1/6} \simeq
10^{-17} \left(\frac{g_{*}}{100} \right)^{1/2} \left[ \frac{g(\tau)}{g_*} \right]^{1/6}
T_{\rm GeV} \, ,
\end{equation}
where $T_{\rm GeV}$ is the numerical value of the temperature in GeV.

We compare the results of the gravitational-wave background in the scenario under
consideration with a number of constraints --- stemming from the BBN prediction of the
abundance of light elements, the CMB temperature anisotropy, and the pulsar timing arrays
--- and with the sensitivity limits of the current and future experiments on direct
gravitational-wave detection (LIGO, LISA, and BBO).

\subsection{The BBN constraint}
\label{sec:bbn}

The energy density $\rho_{\rm rel}$ of relativistic matter other than photons in the
early universe is encoded in the effective number of neutrino species $ N_{\nu} $\,:
\begin{equation} \label{bbn_density}
\frac{\rho_{\rm rel}}{\rho_{\gamma}} = \frac{7}{8} N_{\nu} \, .
\end{equation}
The BBN constraints on the quantity $ N_{\nu} $ are ranging from $ 3.04 $ to about 5
\cite{Maggiore:1999vm}. The measurements of the light-element abundances combined with
the analysis of the Wilkinson Microwave Anisotropy Probe (WMAP) data \cite{Cyburt:2004yc}
give the constraint $ 2.67 < N_{\nu} < 3.65 $ at 68\% CL.

Gravitational waves are among the relativistic components in (\ref{bbn_density}), which
leads to upper bounds on their energy density.  According to (\ref{qH}), the comoving
wave number associated with the Hubble radius at that time is equal to $q_{\rm NS} \simeq
5 \times 10^{-22}$, which is very small.  Thus, we can use equations (\ref{upsilon}) and
(\ref{utot}) to obtain the level of gravitational waves during the epoch of
nucleosynthesis:
\begin{equation} \label{bbn}
\frac{\rho_\gw}{\rho_\gamma} = \frac{\rho_\gw}{\rho_r} \frac{\rho_r}{\rho_\gamma} = r_B^2
\Upsilon_\gw \left( \frac{g}{g_* } \right)^{1/3} \frac{g}{2} \simeq 4.8 \times 10^{-18}
r_B^2 \, ,
\end{equation}
where we have substituted $g = 3.36$ and $g_* = 100$.  For the natural upper bound $r_B <
1$ (in typical scenarios, we have \cite{Boyarsky:2011uy} $ r_B < 10^{-5}$), the BBN
constraint is satisfied with a huge margin.

\subsection{The CMB constraint}

The Cosmic Bakground Explorer (COBE) measurements give the today's energy density
constraint \cite{Maggiore:2000gv}
\begin{equation}
h^2 \Omega_\gw (f) < 7 \times 10^{-11} \left( \frac{H_0}{f} \right)^2
\end{equation}
on gravitational waves in the frequency region $ f \in [3 \times 10^{-18}, 10^{-16}]$~Hz.
In particular, we have
\begin{equation} \label{cobe} \begin{array}{ll}
h^2 \Omega_\gw(f) < 7.4 \times 10^{-14} \, , &f = 10^{-16}~\mbox{Hz} \, , \medskip \\
h^2 \Omega_\gw(f) < 8.2 \times 10^{-11} \, ,  &f = 3 \times 10^{-18}~\mbox{Hz} \, .
\end{array}
\end{equation}

The relevant region of comoving wave numbers, according to (\ref{qtonu}), is $q \in [ 2
\times 10^{-28}, \, 6.5 \times 10^{-27}] $.  Using (\ref{presday}) and (\ref{ufinal}) and
taking into account that today $\Omega_r \simeq 4 \times 10^{-5}$, we obtain
\begin{equation} \begin{array}{ll} \label{gw-num}
h^2 \Omega_\gw(f) \sim 9 \times 10^{-60} r^2_B \, , &f = 10^{-16}~\mbox{Hz} \,
, \medskip \\
h^2 \Omega_\gw(f) \sim 3 \times 10^{-64} r^2_B \, , &f = 3 \times 10^{-18}~\mbox{Hz} \, .
\end{array}
\end{equation}

Again, given the upper limit $r_B < 1$, we see that the COBE constraints (\ref{cobe}) are
satisfied with a large margin.

\subsection{The combined constraint from the CMB and LSS}

The scalar perturbations power spectrum $\Delta^2_R (k ,\tau)$ is very close to
scale-invariant and is time-independent on superhorizon spatial scales, the WMAP
measurements effectively fix the power spectrum of scalar perturbations at the end of
inflation $ \tau_i $ as $\Delta^2_R (k_* ,\tau_i) \simeq 2.43 \times10^{-9}$ at the
comoving wave number $k_* = 0.002$~Mpc$^{-1}$ \cite{arXiv:1001.4538}. The power spectrum
of tensor perturbations is then $ \Delta^2_h (k_* ,\tau_i) = r \Delta^2_R (k_*,\tau_i)$,
where the tensor-to-scalar ratio $r$ has recently been constrained by the BICEP2
collaboration as $ r = 0.2^{+0.07}_{-0.05} $ \cite{Ade:2014xna} (see, however, the
criticism of this result in \cite{Mortonson:2014bja, Flauger:2014qra}). We can estimate
the contribution of primordial gravitational waves to the quantity $ \Omega_\gw(f) $ at
the frequencies corresponding to the horizon size at recombination ($ f \sim
10^{-16}$~Hz) \cite{arXiv:1001.4538}:
\begin{equation}
h^2 \Omega^{\rm prim}_\gw(f) = \frac{1}{12} \left( \frac{2 \pi f}{H_0} \right)^2
\Delta^2_h (q_{\rm cmb},\tau_0) \, ,
\end{equation}
where $q_{\rm cmb}$ is the corresponding wavenumber.

The current power of gravitational waves that crossed the horizon during recombination
can be estimated as follows:
\begin{equation}
\Delta^2_h (q_{\rm cmb},\tau_0) \simeq  \frac{\Delta^2_h (q_{\rm cmb},\tau_{\rm cmb})}{(1
+ z_{\rm cmb})^2} \simeq \frac{r \Delta^2_R (q_{\rm cmb},\tau_i)}{(1 + z_{\rm cmb})^2} \,
.
\end{equation}
The estimate $ r \lesssim 0.2 $ then translates to the level of the gravitational-wave
background
\begin{equation}
h^2 \Omega_\gw(f) = \frac{1}{12} \left( \frac{2 \pi f}{H_0} \right)^2 \frac{r \Delta^2_R
(q_{\rm cmb},\tau_i)}{(1 + z_{\rm cmb})^2} \lesssim 10^{-12}\, .
\end{equation}
Using (\ref{gw-num}) for $ f \sim 10^{-16}$~Hz, we see that this constraint is also very
well satisfied.

\subsection{Constraints from the pulsar timing measurements}

Constraints from pulsar timing measurements are sensitive to the frequency region $ f \in
[10^{-10}, 10^{-8}] $ Hz, which corresponds to $q \in [6.5 \times 10^{-21}, 6.5 \times
10^{-19}]$.  In this domain, our results for the spectral energy density of gravitational
waves produced by helical magnetic field read:
\begin{equation} \label{pulsar} \begin{array}{ll}
h^2 \Omega_\gw(f) \sim 5 \times 10^{- 37} r^2_B \, , &f = 10^{-8}~\mbox{Hz} \, ,
\medskip \\
h^2 \Omega_\gw(f) \sim 9 \times 10^{- 40} r^2_B \, , &f = 10^{-9}~\mbox{Hz} \, ,
\medskip \\
h^2 \Omega_\gw(f) \sim 1.5 \times 10^{- 42} r^2_B \, , &f = 10^{-10}~\mbox{Hz} \, .
\end{array}
\end{equation}

Pulsar-timing experiments have currently placed an upper bound $ \Omega_\gw (f ) < 2
\times 10^{-8}$ at frequencies $ 10^{-9}~{\rm Hz} < f < 10^{-8}~{\rm Hz}$
\cite{Boyle:2007zx}. In the coming years, the Parkes Pulsar Timing Array (PPTA), which is
already operating, should reach a sensitivity of $ \Omega_\gw (f) \sim 10^{-10} $ (with $
h = 0.72 $) or somewhat better at similar frequencies \cite{Boyle:2007zx}. All these
estimates are passed with a huge margin by estimates (\ref{pulsar}) provided $r_B < 1$.

\subsection{Detector constraints}

Gravitational-wave detectors have maximum sensitivities in the range from about $0.3$~mHz
to about 170~Hz, corresponding to $q$ from $2 \times 10^{-14}$ to $10^{-8}$. We will
analyze the today's spectral density parameter $ h^2 \Omega_\gw (f) $ using our estimates
(\ref{presday}) and (\ref{ufinal}).

The LIGO bounds \cite{arXiv:0910.5772}
\begin{equation} \label{ligo}
h^2 \Omega_\gw(f) < 1.3 \times 10^{-5} \, , \qquad 40~{\rm Hz} < f < 170~{\rm Hz} \, ,
\end{equation}
would give the formal constraint $r^2_B \lesssim  10^{18}$, and are therefore well
satisfied for $r_B < 1$.

For the Laser Interferometer Space Antenna (LISA), the lower threshold of detection is $
h^2 \Omega_\gw (1~\mbox{mHz}) > 10^{-12}$ \cite{Maggiore:2000gv}.  This would require an
unreasonable value $ r^2_B \gtrsim 10^{16} $ in order that gravitational waves produced
by magnetic fields in the scenario under consideration could be detected.  The Big Bang
Observer (BBO) sensitivity \cite{Boyle:2007zx} $ h^2 \Omega_\gw (0.3~\mbox{mHz}) >
10^{-17}$ would require $ r^2_B \gtrsim 5 \times 10^{11} $.

As we have already noted, the characteristic expected values for the fraction of the
magnetic-field energy density after the electroweak phase transition is $ r_B \sim
10^{-4}$ \cite{rubakov} or $ r_B \lesssim 10^{-5}$ in the scenario under consideration
\cite{Boyarsky:2011uy}. Thus, we can conclude that the amount of gravitational waves
generated by magnetic fields in the scenario under consideration \cite{Boyarsky:2011uy}
is far beyond any practical detection.

\subsection{Effects of the finiteness of the magnetic-field generation time and of
damping}

In the above estimates, we assumed that magnetic fields was switched on instantaneously
at some moment of time $\tau_\ini$.  In reality, the generation of magnetic field is a
continuous process. Let us, therefore, take into account that helical magnetic field is
monotonically generated during some period of time $\Delta \tau_\ini \lesssim \tau_\ini$.
Then, for modes with $ q \Delta \tau_\ini \lesssim 1 $, i.e., for $q \lesssim 10^{-16}$,
the spectrum will not be much affected due to the logarithmic dependence on $ \tau_\ini $
in equations (\ref{int}) or (\ref{amp}). For wavenumbers $q \Delta \tau_\ini \gg 1$, the
spectrum will only be damped.  Similar conclusions can be made concerning the possible
relatively rapid decay of magnetic fields after the time $\tau_\fin$.

\section{Constraints on the chiral asymmetry}
\label{sec:chirality}

Previously in this paper, we adopted the characteristic value $\Delta \mu_0 = 6 \times
10^{-5}$ for the initial asymmetry in the chemical parameter, with which we have obtained
rather small estimates for the generated gravitational waves. In this section, we would
like to determine the value of $\Delta \mu_0$ for which the generated gravitational waves
in the model under consideration may reach the threshold of detection.

In the region of small wave numbers $q$, given by inequality opposite to (\ref{qmin}),
one can use expression (\ref{ufinal}) for the spectral power $\Upsilon_q$. It is clear
that the power of the gravitational waves grows with decreasing $k_0$, which is directly
proportional to $ \mu_0 $ via (\ref{mu0}). The characteristic wavelength of magnetic
fields produced by casual mechanism is smaller than the size of the causally connected
region, which implies $ k_0 \gtrsim 10^{-16} $, or $ \Delta \mu_0 \gtrsim 10^{-13} $. On
the other hand, by using (\ref{ufinal}), one can show that, in order to have sufficient
power of gravitational waves to influence the CMB, even assuming $r_B = 1$, we need $
\Delta \mu_0 \lesssim 10^{-20} $. Similarly, for the gravitational waves to be detectable
in the pulsar-timing measurements, we require $ \Delta \mu_0 \lesssim
10^{-14}$$-$$10^{-13} $. Therefore, we can conclude that gravitational waves from casual
magnetic fields cannot be detected in the CMB or pulsar-timing measurements.

Thus, we have to look on the constraints coming from the detector experiments and BBN\@.
Using equation (\ref{final}) with $R$ given by (\ref{ration}), we obtain numerical
estimates for the threshold value of $ \Delta \mu_0 $ above which the gravitational waves
can be detected by experiments or can influence BBN\@.  The estimates are performed along
the same lines as in Sec.~\ref{sec:constraints}, only in this case we keep the value of
$r_B$ fixed and vary the value of $\Delta \mu_0$.  Our results are summarized in
Table~\ref{table}.

\begin{table}[ht]
\setlength{\tabcolsep}{5pt}
\renewcommand{\arraystretch}{1.5}
\begin{tabular}{|c|c|c|c|c|c|c|}
\hline
 $ r_B $ & $ \beta $ & LIGO & LISA & BBO & BBN \\ \hline
 1 & $ 0.25 $ & $ (0.90$$ - $$1.10) \times 10^{-4} $ & $ 0.85  \times 10^{-4} $ & $ 0.80  \times 10^{-4} $ & $ 0.98  \times 10^{-4} $ \\ \hline
 $ 10^{-5} $ & $ 0.25 $ & $ (1.0$$ - $$1.19) \times 10^{-4} $ & $ 0.96  \times 10^{-4} $ & $ 0.91  \times 10^{-4} $ & $ 1.07  \times 10^{-4} $ \\ \hline
 1 & $ 0.35 $ & $(1.73$$ - $$2.1) \times 10^{-4} $ & $ 1.65  \times 10^{-4} $ & $ 1.55  \times 10^{-4} $ & $ 1.88  \times 10^{-4} $ \\ \hline
 $ 10^{-5} $ & $ 0.35 $ & $ (1.92$$ - $$2.26) \times 10^{-4} $ & $ 1.85  \times 10^{-4} $ & $ 1.76  \times 10^{-4} $ & $ 2.06  \times 10^{-4} $ \\ \hline
 1 & $ 0.45 $ & $ (3.23$$ - $$3.89) \times 10^{-4} $ & $ 3.10  \times 10^{-4} $ & $ 2.92  \times 10^{-4} $ & $ 3.50  \times 10^{-4} $ \\ \hline
 $ 10^{-5} $ & $ 0.45 $ & $ (3.56$$ - $$4.15) \times 10^{-4} $ & $ 3.4  \times 10^{-4} $ & $ 3.31  \times 10^{-4} $ & $ 3.82  \times 10^{-4} $ \\ \hline
\end{tabular}
\caption{Estimates for the threshold of the parameter $\Delta\mu_0$ above which the
generated gravitational waves can be detected by experiments or can influence BBN.
\label{table}}
\end{table}

\section{Summary and conclusions}
\label{sec:summary}

In this work, we have studied the background of gravitational waves produced by the
dynamics of magnetic fields excited close to the epoch of electroweak phase transition
and subsequently evolved via inverse cascade mechanism proposed in
\cite{Boyarsky:2011uy}. We have found that the level of produced gravitational waves is
mainly determined by the details of its generation, depending on the cosmological time of
production and initial power spectrum.  The result is exponentially sensitive to the
value of the initial (at temperature $T_\ini \approx 100$\,GeV) chiral asymmetry
specified by the conformal difference $\Delta\mu_0$ in the chemical potentials, which is
related to the cutoff scale $k_0$ in the initial magnetic-field power spectrum via
(\ref{mu0}). For the characteristic value $\Delta \mu_0 = 6 \times 10^{-5}$, used in the
scenario of \cite{Boyarsky:2011uy}, the level of gravitational waves is many orders of
magnitude beyond practical detection either indirectly (by its impact on the CMB
temperature anisotropy) or directly. However, an increase to the level of $\Delta \mu_0
\simeq 10^{-4}$ will make the gravitational waves detectable in a number of experiments
and will affect the BBN scenario.  The specific thresholds for $\Delta \mu_0$ are
indicated in Table~\ref{table} for two different values of the initial fraction $r_B$ of
the energy density of magnetic field and for three different values of the power $\beta$
in (\ref{evolu}) that determines the evolution of magnetic field.

\acknowledgments

We are grateful to Alexey Boyarsky and Oleg Ruchayskiy for the statement of the problem
and for valuable comments, and to Maksym Sydorenko for helpful discussions. This work was
supported by the Swiss National Science Foundation grant SCOPE IZ7370-152581. O.~T. is
grateful to the Scientific and Educational Center of the Bogolyubov Institute for
Theoretical Physics for support. The work of O.~T. was supported in part by the WFS
National Scholarship Programme and also by the Deutsche Forschungsgemeinschaft DFG in
part through the Collaborative Research Center ``The Low-Energy Frontier of the Standard
Model" (SFB 1044), in part through the Graduate School ``Symmetry Breaking in Fundamental
Interactions" (DFG/GRK 1581), and in part through the Cluster of Excellence "Precision
Physics, Fundamental Interactions and Structure of Matter" (PRISMA).  The work of Yu.~S.
was supported in part by the SFFR of Ukraine Grant No.~F53.2/028.

\appendix

\section{Generation of gravitational waves}
\label{sec:gwaves}

Tensor perturbations, or gravitational waves, in an expanding universe are described by
the transverse traceless part of metric perturbations.  Writing the perturbed spatially
flat metric in conformal coordinates as $ d s^2 = a^2 \left( \eta_{\mu \nu} + h_{\mu \nu}
\right)dx^{ \mu } dx^{ \nu }$, for tensor perturbations we have $ h_{00} = h_{0i} =
h^i{}_i = 0$, and the usual gauge conditions $\partial_i h^{ij} = 0 $.

The equation for the evolution of gravitational waves is obtained from the
Hilbert--Einstein action \cite{mukhanov,gorbunov2}:
\begin{equation} \label{gravweq}
\ddot h_{ i j } + 2\frac{\dot a}{a} \dot h_{ i j } - \Delta  h_{ i j }  = - 16 \pi G
\Pi_{i j } \, ,
\end{equation}
where $\Pi_{ij}$ is the transverse traceless part of the stress tensor $T_{ij}$ which, in
the Fourier representation, can be obtained by applying the symmetric transverse
projector $ P_{ij} = \delta_{ij} - \hat p_i \hat p_j\,$:
\begin{eqnarray} \label{pi}
\Pi_{ij} (\bq) = \left[ P^m{}_{(i} P^n{}_{j)} - \frac12 P_{ij} P^{mn} \right] T_{mn} \, .
\end{eqnarray}

In the Fourier representation in the comoving space, we can expand the gravitational
perturbations into the polarization components:
\begin{equation} \label{polars}
h_{ij} (\bq, \tau)= \sum_{\sigma=+,\times} h_{\bq, \sigma}(\tau) \epsilon_{ i j } (\bq,
\sigma) \, .
\end{equation}
The real traceless polarization tensors satisfy the conditions
\begin{equation}
\epsilon_{ij} (\bq,\sigma) = \epsilon_{ij} (-\bq,\sigma) \, , \qquad
q^i \epsilon_{ij} (\bq,\sigma) = 0 \, , \qquad \epsilon^{ij} (\bq,\sigma)\, \epsilon_{ij} (\bq,\sigma') =
\delta_{\sigma\sigma'} \, .
\end{equation}

We introduce the standard variable $v_{\bq, \sigma} = a h_{\bq, \sigma}$.  In a
radiation-dominated universe, we have $\ddot a = 0$, so that from (\ref{gravweq}) we then
have
\begin{equation} \label{eq:grav_waves}
\ddot v_{\bq, \sigma} + q^2  v_{\bq, \sigma}  = - 16 \pi G a \epsilon^{ij}(\bq, \sigma)
\Pi_{ij} (\bq) \equiv f_\sigma (\bq, \tau)\, .
\end{equation}
The solution of this equation with zero initial conditions ($ v_{\bq, \sigma} = 0$, $\dot
v_{\bq, \sigma} = 0 $) at $\tau = \tau_\ini$ is
\begin{equation} \label{vq}
v_{\bq, \sigma} = \frac{1}{q}\int^{\tau}_{\tau_{\rm in}} \sin [q (\tau - \tau')]
f_\sigma (\bq,\tau') d \tau' \, .
\end{equation}

The energy density of the modes inside the Hubble radius gives the Fourier representation
of the gravitational-wave background in the form:
\begin{equation} \label{enden}
\rho_{\rm gw} = \frac{1}{64 \pi G a^4} \int \frac{d^3 \bq}{(2 \pi)^3} \sum_\sigma \left(
\dot v_{\bq, \sigma} \dot v_{-\bq, \sigma} + q^2 v_{\bq, \sigma} v_{-\bq, \sigma} \right)
\, ,
\end{equation}
where the integral proceeds over the values of $\bq$ inside the Hubble radius at a given
moment of time.  For free gravitational waves, i.e., with zero right-hand side in
(\ref{eq:grav_waves}), we have
\begin{equation} \label{gw}
\rho_{\rm gw} = \frac{1}{64 \pi G a^4} \int \frac{d^3 \bq}{(2 \pi)^3} q^2 \sum_\sigma
\left| \bar v_{\bq, \sigma} \right|^2 \, ,
\end{equation}
where $\bar v_{\bq, \sigma}$ is the amplitude of the harmonic function $v_{\bq, \sigma}$
with respect to its oscillations in time.  Equation (\ref{gw}) is applicable to the
situation where the source of gravitational waves has already stopped operating.

\section{The power spectrum of gravitational waves}
\label{sec:gravcalc}

From (\ref{eq:grav_waves}), (\ref{vq}) and (\ref{set}), it is clear that the energy
density of gravitational waves $\rho_\gw$, which is given by ensemble averaging of
Eq.~(\ref{gw}), is proportional to the square of the initial amplitude $S_0$ of the
magnetic-field spectral density, defined in (\ref{specinit}).  Hence, it is reasonable to
consider a normalized dimensionless quantity which is independent of this amplitude [see
Eq.~(\ref{upsilon})]:
\begin{equation}
\Upsilon_\gw = \frac{\rho_\gw}{r^2_B \rho_r} \left[ \frac{g_* }{g (\tau)} \right]^{1/3} =
\frac{\Omega_\gw (\tau)}{r^2_B \Omega_r (\tau)} \left[ \frac{g_*}{g (\tau)} \right]^{1/3}
\, .
\end{equation}
For this quantity, by averaging (\ref{gw}) with solution (\ref{vq}) and employing the
Wick property of the magnetic-field statistics, we obtain the following expression:
\begin{eqnarray} \label{upsilonA}
\Upsilon_\gw &=& \frac{3}{8 \pi} \left( \int_0^\infty S(p) {p}^2 d p \right)^{-2} \int
q^4 dq \int d^3 \bk \left( c^4 - 2 c^3 -
c^2 + 2 c + 2 \right) S (k) S (|\bk - \bq|) I_0^2 (\bq, \bk) \nonumber \\
&\equiv& \int \Upsilon_q\, d \ln q \, ,
\end{eqnarray}
so that
\begin{eqnarray} \label{upsilonB}
\Upsilon_q &=& \frac{3 q^5}{8 \pi} \left( \int_0^\infty S(p) {p}^2 d p \right)^{-2} \int
d^3 \bk \left( c^4 - 2 c^3 - c^2 + 2 c + 2 \right) S (k) S (|\bk - \bq|) I_0^2 (\bq, \bk)
\nonumber \\
&=& \frac{8 q^5}{3 \pi^2 S_0^2 k_0^6} \int d^3 \bk \left( c^4 - 2 c^3 - c^2 + 2 c + 2
\right) S (k) S (|\bk - \bq|) I_0^2 (\bq, \bk) \, ,
\end{eqnarray}
where the final result is presented for spectrum (\ref{specinit}).  Here, $c = \hat \bq
\hat \bk = \cos \theta$, and $I_0 (\bq, \bk)$ is the amplitude of the oscillations of the
time integral
\begin{equation} \label{int}
I(\bq, \bk, \tau) \equiv \int^{\tau}_{\tau_{\rm in}} \frac{\sin \left[ q \left( \tau -
\tau' \right) \right]}{q \tau'} g_k \left( \tau' \right) g_{|\bq - \bk|} \left( \tau'
\right) d \tau' \, ,
\end{equation}
after the source is turned off.  This integral arises as a solution (\ref{vq}) of the
metric-perturbation equation.  The amplitude of its oscillations can be expressed as
\begin{equation} \label{amp}
I_0 (\bq, \bk) \equiv \left| \int^{\tau_\fin}_{\tau_{\rm in}} \frac{d \tau}{q \tau} e^{ -
i q \tau} g_k \left( \tau \right) g_{|\bq - \bk|} \left( \tau \right) d \tau \right| \, .
\end{equation}

The main impact to the quantity $\Upsilon_q$, defined in (\ref{upsilonA}) and
(\ref{amp}), comes from the region of $k$ or $|\bk - \bq |$ that maximizes the product of
the growth factor $ g_k $ and the initial spectral power $S (k)$.

\textbf{1.} For sufficiently large values of $q$, we can estimate the time integral in
(\ref{amp}) by using decomposition into slow-varying and rapidly varying functions. To
this purpose, we note that the product $g_k \left( \tau \right) g_{|\bq - \bk|} \left(
\tau \right)$ at time $\tau$ is peaked in momentum space around the argument $k + | \bq -
\bk| \simeq \left( k^2 + |\bq - \bk|^2 \right)^{1/2} \simeq k_m \approx \cB (\tau) / 2
\cA (\tau)$. Regarding this product as a slowly varying function compared to the first
exponent in the integral
\begin{equation} \label{integral}
\int^{\tau_\fin}_{\tau_\ini} e^{- i q \tau - \ln q \tau} \times  e^{- \cA (\tau) ( k^2 +
|\bq - \bk|^2 ) + \cB (\tau) ( k + |\bq - \bk| )  } d \tau \, ,
\end{equation}
we require the following condition to be satisfied at this momentum:
\begin{equation}
\frac{\left| - \dot \cA k_m^2 + \dot \cB k_m \right|}{\left|- i q - \frac{1}{\tau}
\right|} = \frac{\left| \frac{d}{d \tau} \left[ \frac{\cB^2 (\tau)}{4 \cA (\tau)} \right]
\right|}{\left|- i q - \frac{1}{\tau} \right|} \ll 1 \, , \qquad \tau_\ini < \tau <
\tau_\fin \, .
\end{equation}
Using (\ref{AB}) and (\ref{evolu}), we can write this estimate in the form
\begin{equation} \label{adiab}
\frac{9 k_0^2 \tau_\ini}{16 \sigma_c} \frac{F \left( \tau/ \tau_\ini \right)}{\sqrt{ (q
\tau_\ini)^2 + (\tau_\ini / \tau)^2}} \ll 1 \, , \qquad \tau_\ini < \tau < \tau_\fin \, ,
\end{equation}
where
\begin{equation}
F (x) \equiv \frac{d}{dx} \left[ \frac{\left( \int_1^x y^{-\beta} dy \right)^2}{x - 1}
\right] \, .
\end{equation}
Expression (\ref{adiab}) is maximal approximately at $\tau/\tau_\ini \simeq 1/q
\tau_\ini$ if $q \tau_\ini < 1$, where it is estimated as
\begin{equation}
\frac{9 k_0^2}{16 q \sigma_c} \frac{1 - 2 \beta}{(1 - \beta)^2} ( q \tau_\ini)^{2
\beta } \approx \frac{9 k_0^2}{16 q \sigma_c} ( q \tau_\ini)^{2 \beta } \, ,
\end{equation}
and it is maximal and equal to $9 k_0^2 /16 q \sigma_c$ at $\tau = \tau_\ini$ if $q
\tau_\ini > 1$.  Taking all this into account, we obtain the condition of applicability
of this method in the form:
\begin{equation} \label{app:qmin}
q \tau_\ini \gtrsim \left( \frac{9 k_0^2 \tau_\ini}{16 \sigma_c} \right)^{1/(1 - 2
\beta)} \approx 10^{-3} \, ,
\end{equation}
where the numerical estimate is made for the parameter values described in
Sec.~\ref{sec:gw-field}, i.e., $\sigma_c \simeq 70$, $k_0 = 4.6 \times 10^{-8}$,
$\tau_\ini = 7.3 \times 10^{15}$, and $\beta = 0.35$. It is valid then for $q \gtrsim
10^{-19}$. For such values of $q$, integral (\ref{integral}) can be estimated as
\begin{eqnarray} \label{integral1}
\int^{\tau_\fin}_{\tau_\ini} e^{- i q \tau - \ln q \tau} \times  e^{- \cA (\tau) ( k^2 +
|\bq - \bk|^2 ) + \cB (\tau) ( k + |\bq - \bk| )  } d \tau  \nonumber \\
\approx \left. \frac{G (q\tau)}{q} e^{- \cA (\tau) ( k^2 + |\bq - \bk|^2 ) + \cB (\tau) (
k + |\bq - \bk| ) } \right|_{\tau_\ini}^{\tau_\fin} \, ,
\end{eqnarray}
where $G(q \tau)$ is the primitive of the first exponent in (\ref{integral1}) which
oscillates around zero, i.e.,
\begin{equation} \label{primitive}
G (x) = - \int_x^\infty \frac{d y}{y} e^{- i y} = {\rm ci} (x) - i\, {\rm si} (x) \, .
\end{equation}
For this function, we can use the following good approximation:
\begin{equation}
\left| G (x) \right| \simeq \ln \left( 1 + \frac{1}{x} \right) \, .
\end{equation}
Thus, we have the following estimate for amplitude (\ref{amp}) and for $q$ satisfying
(\ref{app:qmin}):
\begin{equation} \label{amp1}
I_0^2 (\bq,\bk)  \simeq \frac{1}{q^2} \left[ \ln^2 \left( 1 + \frac{1}{q \tau_\ini}
\right) + \ln^2 \left( 1 + \frac{1}{q \tau_\fin} \right) e^{- 2 \cA_\fin ( k^2 + |\bq -
\bk|^2 ) + 2 \cB_\fin ( k + |\bq - \bk| ) } \right] \, ,
\end{equation}
where $\cA_\fin = \cA (\tau_\fin)$ and $\cB_\fin = \cB (\tau_\fin)$.

The peak in the momentum distribution in the exponent of (\ref{amp1}) is located around
\begin{equation} \label{km}
k_m \approx \frac{\cB_\fin}{2 \cA_\fin} = \frac{3}{4 (1 - \beta)} \left(
\frac{\tau_\ini}{\tau_\fin} \right)^\beta k_0 \approx \left( \frac{\tau_\ini}{\tau_\fin}
\right)^\beta k_0 \, .
\end{equation}
The dispersion of the momentum distribution is given by $\Delta k \simeq 1 / \sqrt{4
\cA_\fin} \simeq \sqrt{\sigma_c / 4  \tau_\fin}$.  One can check that $\Delta k / k_m \ll
1$ as long as $\tau_\fin/ \tau_\ini \gtrsim 10^2$. This is true in our case, and the
momentum distribution in (\ref{amp1}) is thus reasonably narrow.  It is also
exponentially highly peaked, so one should estimate its contribution in integral
(\ref{upsilonB}) relative to the contribution of unity in the brackets of (\ref{amp1}).
This is relevant only for the case
\begin{equation}
q \lesssim k_m \approx \left( \frac{\tau_\ini}{\tau_\fin} \right)^\beta k_0 \, ,
\end{equation}
where we can neglect the vector $\bq$ in the exponent of (\ref{amp1}) and obtain for the
ratio of these contributions:
\begin{eqnarray} \label{app:ration}
R = \frac{\displaystyle \ln^2 \left(1 + \frac{1}{q \tau_\fin} \right) e^{\cB_\fin^2/
\cA_\fin} \int_0^\infty dk k^{6} e^{- 4 \cA_\fin (k - k_m)^2}} {\displaystyle \ln^2
\left(1 + \frac{1}{q \tau_\ini} \right) \int_0^\infty dk k^{6} e^{- 2 k^2 / k_0^2}} \sim
\frac{6\, e^{\cB_\fin^2/ \cA_\fin} }{\sqrt{\cA_\fin}\, k_0 } \left(
\frac{\tau_\ini}{\tau_\fin} \right)^{2 + 6 \beta} \nonumber \\
= \frac{6}{k_0} \sqrt{\frac{\sigma_c}{\tau_\fin}} \left(\frac{\tau_\ini}{\tau_\fin}
\right)^{2 + 6 \beta} \exp \left[{\frac{9 k_0^2 \tau_\fin}{4 (1 - \beta)^2 \sigma_c}
\left( \frac{\tau_\ini}{\tau_\fin} \right)^{2 \beta}} \right]  \, .
\end{eqnarray}
We have $R  \sim 10^{-9}$ for $k_0 = 4.6 \times 10^{-8}$, $\tau_\ini = 7.3 \times
10^{15}$, and $\tau_\fin = 1.5 \times 10^{19}$, assuming also $q \tau_\ini \gtrsim 1$.
Although this number is small for the scenario of \cite{Boyarsky:2011uy}, one can see
that the expression depends exponentially on the initial and final times, $\tau_\ini$ and
$\tau_\fin$, and on the value of $k_0$, and potentially might become large in other
scenarios of magnetic-field generation. If this is the case, and the exponent in
(\ref{amp1}) dominates, then, by calculating (\ref{upsilonB}) with spectrum
(\ref{specinit}), we get the result
\begin{eqnarray} \label{inter}
\Upsilon_q &\simeq& \frac{6 q^3}{k_0^4} \ln^2 \left(1 + \frac{1}{q \tau_\fin} \right)
\frac{ e^{\cB_\fin^2/ \cA_\fin}}{\sqrt{\cA_\fin} } \left(\frac{\tau_\ini}{\tau_\fin}
\right)^{6 \beta} \nonumber \\ &\simeq& \frac{6 q^3}{k_0^4} \ln^2 \left(1 + \frac{1}{q
\tau_\fin} \right) \sqrt{\frac{\sigma_c}{\tau_\fin}} \left(\frac{\tau_\ini}{\tau_\fin}
\right)^{6\beta} \exp \left[{\frac{9 k_0^2 \tau_\fin}{4 (1 - \beta)^2 \sigma_c} \left(
\frac{\tau_\ini}{\tau_\fin} \right)^{2\beta}} \right] \, .
\end{eqnarray}
In the range $k_m \lesssim q \lesssim k_0$, the factor $S(| \bk - \bq|)$ in
(\ref{upsilonB}) and the exponent in the brackets of (\ref{amp1}) produce an additional
exponential suppression, leading to multiplication of the result (\ref{inter}) by the
factor $\left( q^2 / k_m^2 \right) \exp \left( - 2 A_\fin q^2 \right) = \left(
{\tau_\fin} / {\tau_\ini} \right)^{2 \beta} \left( q^2 / k_0^2 \right) \exp \left( - 2
\tau_\fin q^2 / \sigma_c \right)$.

These results can be combined together as
\begin{equation} \label{final1}
\Upsilon_q \simeq R \left(\frac{q}{k_0} \right)^3 \ln^2 \left( 1 + \frac{1}{q \tau_\fin}
\right) \left[ 1 + \left( \frac{\tau_\fin}{\tau_\ini} \right)^{2 \beta} \frac{q^2}{k_0^2}
\right] e^{- 2 \cA_\fin q^2}  \, ,
\end{equation}
where $R$ is given by (\ref{app:ration}).

In the case where ratio (\ref{app:ration}) is small, $R \ll 1$, our estimate of
(\ref{upsilonB}) becomes
\begin{equation} \label{nosup}
\Upsilon_q \simeq \left(\frac{q}{k_0} \right)^3 \ln^2 \left( 1 + \frac{1}{q \tau_\ini}
\right) \, , \qquad q \lesssim k_0 \, .
\end{equation}
For $q \gtrsim k_0$, we get an extra suppression in (\ref{upsilonB}) coming from the
initial power spectrum (\ref{specinit}), with the result
\begin{equation} \label{sup}
\Upsilon_q \simeq \left(\frac{q}{k_0} \right)^5 \ln^2 \left( 1 + \frac{1}{q \tau_\ini}
\right) e^{-q^2/k_0^2} \, .
\end{equation}
This result of our estimates of the quantity $\Upsilon_q$ is presented in
Fig.~\ref{fig:yq}.

Equations (\ref{final1})--(\ref{sup}) can be combined to give the final approximation in
the form
\begin{eqnarray} \label{app:final}
\Upsilon_q &\simeq& \left(\frac{q}{k_0} \right)^3 \left[ \ln^2 \left( 1 + \frac{1}{q
\tau_\ini} \right) \left( 1 + \frac{q^2}{k_0^2} \right) \phantom{\left(
\frac{\tau_\fin}{\tau_\ini} \right)^{2 \beta}} \right. \nonumber\\
&&\left. {} + R \ln^2 \left( 1 + \frac{1}{q \tau_\fin} \right) \left( 1 + \left(
\frac{\tau_\fin}{\tau_\ini} \right)^{2 \beta} \frac{q^2}{k_0^2} \right) e^{- 2 \cA_\fin
q^2} \right] e^{-q^2/k_0^2} \, ,
\end{eqnarray}
where $R$ is given by (\ref{app:ration}).

{\bf 2.} For small values of $q$, in the case opposite to (\ref{app:qmin}), we can
neglect $q$ with respect to $k$ under the integrals in expressions (\ref{upsilonB}) and
(\ref{amp}) to write
\begin{eqnarray} \label{upsilonC}
\Upsilon_q &\simeq& \frac{8 q^5}{3 \pi^2 S_0^2 k_0^6} \int d^3 \bk \left( c^4 - 2 c^3 -
c^2 + 2 c + 2 \right) S^2 (k) I_0^2 (q, k) \nonumber \\
&\simeq& \frac{6 q^5}{S_0^2 k_0^6} \int S^2 (k) I_0^2 (q, k) k^2 d k \, ,
\end{eqnarray}
\begin{equation} \label{ampC}
I_0 (q, k) \simeq \left| \int^{\tau_\fin}_{\tau_{\rm in}} \frac{d \tau}{q \tau} e^{ - i q
\tau} e^{- 2 \cA (\tau) k^2 + 2 \cB (\tau) k } d \tau \right| \, .
\end{equation}

The quantity
\begin{eqnarray}
I_0^2 (q, k) \simeq \frac{1}{q^2}  \int^{\tau_\fin}_{\tau_{\rm in}}
\int^{\tau_\fin}_{\tau_{\rm in}} \frac{d \tau d \tau'}{ \tau \tau'} \cos( q (\tau-\tau'))
e^{- 2 (\cA (\tau)+\cA (\tau')) k^2 + 2 (\cB (\tau)+\cB (\tau')) k } d \tau d \tau' \, ,
\end{eqnarray}
for small values of $q \lesssim 1/\tau_\fin$, simplifies to
\begin{eqnarray}
I_0^2 (q, k) \simeq \frac{1}{q^2}  \int^{\tau_\fin}_{\tau_{\rm in}}
\int^{\tau_\fin}_{\tau_{\rm in}} \frac{d \tau d \tau'}{ \tau \tau'}  e^{- 2 (\cA
(\tau)+\cA (\tau')) k^2 + 2 (\cB (\tau)+\cB (\tau')) k } d \tau d \tau'  \, ,
\end{eqnarray}
and the $q$-dependence of the spectrum of gravitational waves is given by
\begin{eqnarray} \label{upsilon_small}
\Upsilon_q &\simeq& \frac{q^3}{ k_0^3} f(\tau_{\rm in}, \tau_\fin) \, ,
\end{eqnarray}
For the values $\tau_\ini = 7.3 \times 10^{15}$ and $\tau_\fin = 1.5 \times 10^{19}$, the
prefactor $ f(\tau_{\rm in}, \tau_\fin) $ is approximately equal to $5$.

The results for small and large values of $q$ can be combined together by an
extrapolation. In the case of $R \ll 1$, equation (\ref{app:final}) will be a reasonable
extrapolation, so that we adopt, finally,
\begin{equation} \label{app:ufinal}
\Upsilon_q \simeq \left(\frac{q}{k_0} \right)^3 \left( 1 + \frac{q^2}{k_0^2} \right)
\ln^2 \left( 1 + \frac{1}{q \tau_\ini} \right) e^{-q^2/k_0^2} \, .
\end{equation}

\section{Monochromatic source of magnetic field}
\label{sec:monosource}

In this section, following \cite{Boyarsky:2011uy}, we consider a simple case of
monochromatic magnetic field which may be useful for understanding the dependence of the
expected power of gravitational waves on the initial spectrum of magnetic field. The
initial spectrum in this case is approximated by the isotropic form
\begin{equation} \label{monos}
S(k) = S_0 k_0 \delta (k - k_0) \, ,
\end{equation}
where the factor $k_0$ serves to bring the canonical dimension of $S_0$ to its dimension
in (\ref{specinit}) independently of the choice of the dimension for the wave number $k$.
The spectrum is characterized by two numbers: the comoving spatial scale $k_0$ and the
amplitude $S_0$. The wave number $k_0$ in this case is related to the value of the
difference of chemical potentials as \cite{Boyarsky:2011uy}
\begin{equation}
\Delta \mu = \frac{2 \pi k_0}{\alpha} \, .
\end{equation}

In the case under consideration, we have $g_{k_0} (\tau) \equiv 1$ for the growth factor,
which simplifies the analysis.  The quantity $I_0(\bq, \bk)$ is given by the expression
\begin{equation} \label{int0}
I_0(\bq, \bk) \equiv I (q) = \left| \int^{\tau_\fin}_{\tau_{\rm in}} \frac{d \tau}{q
\tau} e^{ - i q \tau } d \tau \right| \simeq \frac{1}{q} \log \left( 1 + \frac{1}{q
\tau_\ini} \right) \, .
\end{equation}
The spectral quantity $\Upsilon_q$, defined in (\ref{upsilonA}) is then calculated as
follows:
\begin{eqnarray} \label{mono}
\Upsilon_q &=& \frac{3 q^5 I^2 (q)}{8 \pi} \left( \int_0^\infty S(p) {p}^2 d p
\right)^{-2} \int d^3 \bk \left( c^4 - 2 c^3 - c^2 + 2 c + 2 \right) S (k) S (|\bk -
\bq|) \nonumber \\
&=& \frac{3 q^5 I^2 (q)}{8 \pi k_0^4} \int d^3 \bk \left( c^4 - 2 c^3 - c^2 + 2 c + 2
\right) \delta (k - k_0) \delta (|\bk - \bq| - k_0) \nonumber \\
&=& \frac{3 q^5 I^2 (q)}{4 k_0^4} \int_{-1}^1 dc \left( c^4 - 2 c^3 - c^2 + 2 c + 2
\right) \int k^2 d k\, \delta (k - k_0) \delta \left(
\sqrt{k^2 - 2 k q c + q^2} - k_0 \right) \nonumber \\
&=& \frac{3 q^5 I^2 (q)}{4 k_0^2} \int_{-1}^1 dc \left( c^4 - 2 c^3 - c^2 + 2 c + 2
\right) \delta \left( \sqrt{k_0^2 - 2 k_0 q c + q^2} - k_0 \right) \nonumber \\
&=& \frac{3 q^4 I^2 (q)}{4 k_0^2} \left( c_0^4 - 2 c_0^3 - c_0^2 + 2 c_0 + 2 \right)
\theta(2 k_0 - q) \, ,
\end{eqnarray}
where $c = \hat \bq \hat \bk = \cos \theta$ and $0 < c_0 = q / 2 k_0 \leq 1$.  In this
interval, we have
\begin{equation}
2 < \left( c_0^4 - 2 c_0^3 - c_0^2 + 2 c_0 + 2 \right) < \frac{41}{16} \approx 2.56 \, ,
\end{equation}
so, for an estimate, we can replace this expression by 2 and get
\begin{equation} \label{ups-mono}
\Upsilon_q \approx \frac{3 q^4 I^2 (q)}{2 k_0^2} \theta(2 k_0 - q) \approx \frac{3 q^2}{2
k_0^2} \log^2 \left( 1 + \frac{1}{q \tau_\ini} \right) \theta(2 k_0 - q) \, .
\end{equation}

Note that the gravitational-wave spectrum (\ref{ups-mono}) differs from the spectrum
(\ref{app:ufinal}) by an extra factor $k_0 / q$.  This factor is a specific artefact of
the delta-functional (or sharply-peaked) power spectrum, and can be traced in the
derivation of (\ref{mono}).

By integrating (\ref{upsilon1}) using (\ref{ups-mono}), we get
\begin{equation} \label{utot-mono}
\Upsilon_\gw \simeq \frac{1}{k_0^2 \tau_\ini^2} \left( 1 + \log k_0 \tau_\ini \right) \,
, \quad k_0 \tau_\ini \gg 1 \, .
\end{equation}
This can be compared to (\ref{utot}).  It is clear that the larger is the comoving
spatial scale of the magnetic field (the smaller is $k_0$), the larger is the level of
generated gravitational waves, provided the quantity $r_B$ is kept fixed.

Using (\ref{ups-mono}), we obtain the present-day energy density of gravitational waves
in the monochromatic case under consideration:
\begin{equation} \label{con-mono}
\Omega_\gw (f) = \Omega_r \frac{3 r_B^2}{2 k_0^2 \tau_\ini^2} \left( \frac{g_0}{g_*}
\right)^{1/3} \left( q \tau_\ini \right)^2 \log^2 \left( 1 + \frac{1}{q \tau_\ini}
\right) \theta \left( 2 k_0 - q \right) \, ,
\end{equation}
where the physical frequency $f$ and the dimensionless wave number $q$ are related by
Eq.~(\ref{qtonu}).

Let us compare the results obtained in the case of monochromatic magnetic field with
observational and experimental constraints.

Consider the case characterized by our typical length scale: $k_0 = 10^{-8}$. The level
of gravitational waves during nucleosynthesis in the present case will be larger than
(\ref{bbn}) only by the factor $\left( 1 + \log k_0 \tau_\ini \right) \simeq 20$.

For the CMB constraints \cite{Maggiore:2000gv}, we have $q \tau_\ini \sim 10^{-11}$, and
equation (\ref{con-mono}) gives the upper estimate
\begin{equation} \label{cmb-c}
r_B^2 \lesssim 4 \times 10^{9} k_0^2 \tau_\ini^2 \, ,
\end{equation}
a very weak constraint again. For our characteristic wavenumber $k_0 \simeq 5 \times
10^{-8}$ and $\tau_\ini \simeq 2 \times 10^{16}$, we have  $k_0 \tau_\ini \simeq 10^9$,
so that bound (\ref{cmb-c}) is well satisfied for $r_B < 1$.

For the pulsar-timing measurements \cite{Boyle:2007zx}, we have $q \tau_\ini \sim
10^{-3}$, whence we obtain
\begin{equation}
r_B^2 \lesssim k_0^2 \tau_\ini^2 \, .
\end{equation}

For the LIGO bounds at frequency 170~Hz \cite{arXiv:0910.5772}, we have $q \simeq
10^{-8}$, and the constraint reads
\begin{equation} \label{mono-ligo}
r_B^2 \lesssim  k_0^2 \tau_\ini^2 \, .
\end{equation}

For the detection by the LISA experiment \cite{Maggiore:2000gv}, we need at least
\begin{equation}
r_B^2 \gtrsim 10^{-7} k_0^2 \tau_\ini^2  \, .
\end{equation}

The LIGO constraint (\ref{mono-ligo}) is in agreement with that of
Sec.~\ref{sec:constraints} for our typical values of $k_0 \simeq 5 \times 10^{-8}$ and
$\tau_\ini \simeq 2 \times 10^{16}$. Other constraints in this section are stronger than
those of Sec.~\ref{sec:constraints} because of the small extra phase-space factor $q /
k_0$ in (\ref{app:ufinal}), but still are very weak.


\begin{thebibliography}{99}

\bibitem{Bernet:2008qp}
  M.~L.~Bernet, F.~Miniati, S.~J.~Lilly, P.~P.~Kronberg and M.~Dessauges-Zavadsky,
  \emph{Strong magnetic fields in normal galaxies at high redshift},
  \emph{Nature\/} {\bf 454} (2008) 302
  [arXiv:0807.3347 [astro-ph]].

\bibitem{Wolfe:2008nk}
  A.~M.~Wolfe, R.~A.~Jorgenson, T.~Robishaw, C.~Heiles and J.~X.~Prochaska,
  \emph{An 84-$\mu$G magnetic field in a galaxy at redshift $z = 0.692$},
  \emph{Nature\/} {\bf 455} (2008) 638
  [arXiv:0811.2408 [astro-ph]].

\bibitem{Tavecchio:2010mk}
  F.~Tavecchio, G.~Ghisellini, L.~Foschini, G.~Bonnoli, G.~Ghirlanda and P.~Coppi,
  \emph{The intergalactic magnetic field constrained by Fermi/Large Area Telescope
  observations of the TeV blazar 1ES 0229+200},
  \emph{Mon. Not. Roy. Astron. Soc.}  {\bf 406} (2010) L70
  [arXiv:1004.1329 [astro-ph.CO]].

\bibitem{Ando:2010rb}
  S.~'i.~Ando and A.~Kusenko,
  \emph{Evidence for gamma-ray halos around active galactic nuclei and the first
  measurement of intergalactic magnetic fields},
  \emph{Astrophys. J.}  {\bf 722} (2010) L39
  [arXiv:1005.1924 [astro-ph.HE]].

\bibitem{Neronov:1900zz}
  A.~Neronov and I.~Vovk,
  \emph{Evidence for strong extragalactic magnetic fields from Fermi observations of TeV blazars},
  \emph{Science\/} {\bf 328} (2010) 73
  [arXiv:1006.3504 [astro-ph.HE]].

\bibitem{Widrow:2002ud}
  L.~M.~Widrow,
  \emph{Origin of galactic and extragalactic magnetic fields},
  \emph{Rev. Mod. Phys.}  {\bf 74} (2002) 775
  [astro-ph/0207240].

\bibitem{Kandus:2010nw}
  A.~Kandus, K.~E.~Kunze and C.~G.~Tsagas,
  \emph{Primordial magnetogenesis},
  \emph{Phys. Rept.}  {\bf 505} (2011) 1
  [arXiv:1007.3891 [astro-ph.CO]].

\bibitem{Joyce:1997uy}
  M.~Joyce and M.~E.~Shaposhnikov,
  \emph{Primordial magnetic fields, right electrons, and the Abelian anomaly},
  \emph{Phys. Rev. Lett.}  {\bf 79} (1997) 1193
  [astro-ph/9703005].

\bibitem{Boyarsky:2011uy}
  A.~Boyarsky, J.~Frohlich and O.~Ruchayskiy,
  \emph{Self-consistent evolution of magnetic fields and chiral asymmetry in the early universe},
  \emph{Phys. Rev. Lett.}  {\bf 108} (2012) 031301
  [arXiv:1109.3350 [astro-ph.CO]].

\bibitem{llvol8}
L.~D.~Landau and E.~M.~Lifshitz, \emph{Course of Theoretical Physics, Vol.~8:
Electrodynamics of Continuous Media}, Pergamon Press, Oxford (1984).

\bibitem{arXiv:0909.0622}
C.~Caprini, R.~Durrer and G.~Servant, \emph{The stochastic gravitational wave background
from turbulence and magnetic fields generated by a first-order phase transition},
\emph{JCAP\/} {\bf 12} (2009) 024 [arXiv:astro-ph/0909.0622 [astro-ph.CO]].

\bibitem{Tashiro:2012mf}
H.~Tashiro, T.~Vachaspati and A.~Vilenkin,
  \emph{Chiral effects and cosmic magnetic fields},
  \emph{Phys. Rev. D} {\bf 86} (2012) 105033
  [arXiv:1206.5549 [astro-ph.CO]].

\bibitem{Caprini:2003vc}
  C.~Caprini, R.~Durrer and T.~Kahniashvili,
  \emph{Cosmic microwave background and helical magnetic fields: The tensor mode},
  \emph{Phys. Rev. D} {\bf 69} (2004) 063006
  [arXiv:astro-ph/0304556].

\bibitem{Baym:1997gq}
  G.~Baym and H.~Heiselberg,
  \emph{Electrical conductivity in the early universe},
  \emph{Phys. Rev. D} {\bf 56} (1997) 5254
  [arXiv:astro-ph/9704214].

\bibitem{Maggiore:1999vm}
  M.~Maggiore,
  \emph{Gravitational wave experiments and early universe cosmology},
  \emph{Phys. Rept.}  {\bf 331} (2000) 283
  [arXiv:gr-qc/9909001].

\bibitem{Cyburt:2004yc}
  R.~H.~Cyburt, B.~D.~Fields, K.~A.~Olive and E.~Skillman,
  \emph{New BBN limits on physics beyond the standard model from $^4$He},
  \emph{Astropart. Phys.}  {\bf 23} (2005) 313
  [arXiv:astro-ph/0408033].

\bibitem{Maggiore:2000gv}
  M.~Maggiore,
  \emph{Stochastic backgrounds of gravitational waves},
  arXiv:gr-qc/0008027.

\bibitem{arXiv:1001.4538}
E.~Komatsu {\it et al.} [WMAP Collaboration], \emph{Seven-Year Wilkinson Microwave
Anisotropy Probe (WMAP) Observations: Cosmological Interpretation}, \emph{Astrophys. J.
Suppl.} {\bf 192} (2011) 18 [arXiv:1001.4538 [astro-ph.CO]].

\bibitem{Ade:2014xna}
  P.~A.~R.~Ade {\it et al.}  [BICEP2 Collaboration],
  \emph{Detection of B-Mode Polarization at Degree Angular Scales by BICEP2},
  \emph{Phys. Rev. Lett.}  {\bf 112} (2014) 241101
  [arXiv:1403.3985 [astro-ph.CO]].

\bibitem{Mortonson:2014bja}
  M.~J.~Mortonson and U.~Seljak,
  \emph{A joint analysis of Planck and BICEP2 B modes including dust polarization uncertainty},
  arXiv:1405.5857 [astro-ph.CO].

\bibitem{Flauger:2014qra}
  R.~Flauger, J.~C.~Hill and D.~N.~Spergel,
  \emph{Toward an understanding of foreground emission in the BICEP2 region},
  \emph{JCAP\/} {\bf 1408} (2014) 039
  [arXiv:1405.7351 [astro-ph.CO]].

\bibitem{Boyle:2007zx}
  L.~A.~Boyle and A.~Buonanno,
  \emph{Relating gravitational wave constraints from primordial nucleosynthesis, pulsar timing,
  laser interferometers, and the CMB: Implications for the early universe},
  \emph{Phys. Rev. D} {\bf 78} (2008) 043531
  [arXiv:0708.2279 [astro-ph]].

\bibitem{arXiv:0910.5772}
B.~P.~Abbott {\it et al.} [LIGO Scientific and VIRGO Collaborations], \emph{An upper
limit on the stochastic gravitational-wave background of cosmological origin},
\emph{Nature\/} {\bf 460} (2009) 990 [arXiv:0910.5772 [astro-ph]].

\bibitem{rubakov}
D.~V.~Deryagin, D.~Yu.~Grigoriev, V.~A.~Rubakov and M.~V.~Sazhin, \emph{Possible
anisotropic phases in the early universe and gravitational waves background}, \emph{Mod.
Phys. Lett. A} {\bf 1} (1986) 593.

\bibitem{mukhanov}
V.~Mukhanov, \emph{Physical Foundations of Cosmology}, Cambridge University Press,
Cambridge (2005).

\bibitem{gorbunov2}
D.~S.~Gorbunov and V.~A.~Rubakov, \emph{Introduction to the Theory of the Early Universe:
Cosmological Perturbations and Inflationary Theory}, World Scientific, Singapore (2011).

\end{thebibliography}
\end{document}